\documentstyle{article}
\title{\LARGE Solution of symmetry equation and hierarchy of self dual Yang-Mills systems.}
\author{A.~N.~Leznov\thanks{ Universidad Autonoma del Estado de Morelos, CCICAp,Cuernavaca, Mexico}} \date{}

\newcommand{\rig}[2]{\stackrel{#2\rightarrow}{#1}}

\begin{document}
\maketitle

\maketitle

\begin{abstract}
The solution of symmetry equation for Yang-Mills self dual system is found in explicit form of its raising Hamiltonian operator. Thus explicit form of equations of self dual Yang-Mills hierarchy is constructed. 
\end{abstract}

\section{Introduction}

With each system of differential equations for vector function $u$ is connected its own system of symmetry equations, which arises after formal differentiation of the initial system with respect some independent parameter $u'$ and identification $u'=U, (u_x)'=U_x$ and so on as new unknown function. In few words and very roughly connection between symmetry equation and integrable properties of the initial system may be described as follows. If symmetry equation may be resolve exactly on the class of solution of the initial system, then the initial system may be resolved exactly (in explicit or implicit form) in the sence of Cauche-Kovalvskay. If the symmetry equation posses infinite series of partial solutions then initial system posses infinite series soliton like solutions depending on definite number of arbitrary numerical parameters. Such system are known as integrable one \cite{OVS}.    

The goal f the present paper is to show that self dual Yang-Mills system is integrable one 
(in the sense above) and construct hierarchy of equations of the same notice -SDYM. All equations of this hierarchy are invariant with respect to discrete transformation of self dual system \cite{L5}.

We use the following form for self-dual system \cite{Lez 88}
\begin{equation}
G_{\bar z}G^{-1}=f_y,\quad G_{\bar y}G^{-1}=-f_z,\quad G^{-1}G_z=\tilde
f_{\bar y},\quad G^{-1}G_y=-\tilde f_{\bar z} \label{SD}
\end{equation}
where group-valued function $G$ and algebra-valued functions $f,\tilde f$
take values in some semi simple group (algebra).

Exclusion of the algebra-valued functions $f,\tilde f$ leads to self-dual
system in Yangs form \cite{Y}
\begin{equation}
(G^{-1}G_z)_{\bar z}+(G^{-1}G_y)_{\bar y}=0\quad (G_z G^{-1})_{\bar z}+
(G_y G^{-1})_{\bar y}=0 \label{SDI}
\end{equation}
Equation of second order for algebra-valued functions $f, \tilde f$
arises as a corollary of Mauerer-Cartan identities are as follows \cite{Lez 88}
\begin{equation}
f_{\bar y,y}+f_{\bar z,z}=[f_y,f_z],\quad \tilde f_{\bar y,y}+\tilde
f_{\bar z,z}=[\tilde f_{\bar z},\tilde f_{\bar y}] \label{SDII}
\end{equation}

In what follows we will have a deal with the left equation (\ref{SDII}). This equation is Lagrangian one with $L$ function of the form \cite{Lez 88}
\begin{equation}
L=\int d y d z d \bar y d \bar z ({1\over 2} Sp (f_y f_{\bar y}+f_z f_{\bar z})+{1\over 3}(f[f_y,f_z]))\label{L}
\end{equation}
In construction of Hamiltonian equation via (\ref{L}) we will consider $\bar y$ or $\bar z$ as a evolution variable and thus corresponding canonical impulses $\pi=\bar f_y=\frac{\delta L}{\delta f_{\bar y}}$ and thus corresponding Hamiltonian functions must be constructed by Dirack method of Lagrangian factors and has the form
$$
H=-\int d y d z d \bar y d \bar z ({1\over 2} Sp f_z f_{\bar z}+{1\over 3}(f[f_y,f_z]))+
Sp (\lambda (\pi-\bar f))
$$
The symmetry equation corresponding to (\ref{SDII}) by the rules described above $f'=F$ in the case of self dual system has the form
\begin{equation}
F_{\bar y,y}+F_{\bar z,z}=[F_y,f_z]+[f_y,F_z] \label{S}
\end{equation}
With each solution of symmetry equation $F$ it is possible connect evolution like integrable system
$$
\dot f=F
$$
The infinite number of such integrable systems is self dual Yang-Mills hierarchy.

In what follows we present the realization of described above program in explicit form in the case of arbitrary semi simple algebra. We describe procedure how from initial given solution of the symmetry equation some new one. Usually such procedure is noticed as raising Hamiltonian operator. We conserve this terminology. 

\section{The main assertion}

Let algebra valued function $F$ is solution of the symmetry equation (\ref{S}).
Then function $\tilde F$ defined by its derivatives
\begin{equation}
\tilde F_y=-F_{\bar z}+[f_y,F],\quad \tilde F_z=F_{\bar y}+[f_z,F] \label{MN}
\end{equation}
is solution of the same equation. Or up to unessential function of two variables $\bar y,\bar z$
$$
\tilde F=\int (d y(-F_{\bar z}+[f_y,F])+d z(F_{\bar y}+[f_z,F]))
$$
Integral to be taken on arbitrary contour in $y,z$ plane between two points arbitrary initial one $y_0,z_0$ and the point of observation $y,z$. Integral does not depend on the form of the contour of integration but only from initial and final points.

The proof of this assertion is very simple. From equality of the second mixed derivatives of
$\tilde F$ function it follows immediately that $F$ is solution of symmetry equation (\ref{S}). 
Differentiating the first equation of (\ref{MN}) on $\bar y$ variable the second one on $\bar z$
and taking sum of these results we obtain
$$
\tilde F_{\bar y,y}+\tilde F_{\bar z,z}=[(f_{\bar y,y}+f_{\bar z,z}),F]+[f_y,F_{\bar y}]+[f_z,F_{\bar z}
$$
Substituting in the last expression derivatives of $F$ function with respect to $\bar y,\bar z$ arguments via derivatives of $\tilde F$ function and taking into account equations for $f$ function (\ref{SDII}) we conclude that $\tilde F$ is the solution of symmetry equation (\ref{S}).   

\section{Initial conditions}

The symmetry equation contain only derivatives of $F$ function and thus it has obvious  solution $F=\alpha=Const$. The number of arbitrary parameters coincides with the dimension of semi simple algebra $N$. Except of this the derivatives $f$ with respect to all coordinates are also solution of the symmetry equation. Thus 
$$ 
F_0=\alpha+f_{\bar y}+f_{\bar z}+f_y+f_z
$$
is solution of symmetry equation depending on $N+4$ arbitrary numerical parameters.
The equation for $f$ in usual form (\ref{SDII}) arises after choose $F=f_z$. The second equation of (\ref{MN}) lead to $\tilde F=f_{\bar y}$ and the first one to (\ref{SDII}). 

\section{Discrete transformation of self-dual Yang-Mills system}

Now we would like to show that all systems of self dual Yang-Mills hierarchy are invariant
with respect discrete transformation self dual Yang Mills system \cite{L5}.
Discrete transformation is connected with the choose of the maximal root of semi simple algebra and will be called as transformation of the maximal root. We remind shortly content of \cite{L5}.
 Keeping in mind (\ref{SD}) it is not difficult to show that the following
system of equations for a group-valued element $S$ is self-consistent
(equalities of Maurer-Cartan are satisfied)
\begin{equation}
S^{-1}S_y={1\over f_-}[X^+_M,f_y] +({1\over f_-})_{\bar z}X^+_M,\quad
S^{-1}S_z={1\over f_-}[X^+_M,f_z] -({1\over f_-})_{\bar y}X^+_M \label{SS}
\end{equation}
where $X^+_M$ maximal positive root of the algebra, $f_-$ - coefficient
on $X^-_M$ in decomposition $f$ function on the root system of the algebra.

The algebra-valued function $\rig{f}{M}$, satisfying the self-consistent conditions
(in the sense of equality of the second mixed derivatives)
\begin{equation}
\rig{f}{M}_y=S f_y S^{-1}+S_{\bar z}S^{-1}, \quad \rig{f}{M}_z=S f_z S^{-1}-S_{\bar y}S^{-1}
\label{DT}
\end{equation}
also satisfy the self-dual Yang-Mills equations in the form (\ref{SDII}) and
new group value function $\tilde G=SG$ is solution of the same system in
the Yang's form (\ref{SDI}). 
Now after differentiation (\ref{DT}) on arbitrary parameter and identification $f'=F$ we come to the law of transformation of solution of symmetry equation in a form
$$
\rig{F}{M}_y=S (F_y+[Q,f_y]+Q_{\bar z})S^{-1},\quad \rig{F}{M}_z=S (F_z+[Q,f_z]-Q_{\bar y})S^{-1} 
$$
where $Q=S^{-1}S'$ is some linear combination constructed from component $F$ function.
Now let us differentiate (\ref{DT}) with respect to parameter of evolution. We have
$$
\rig{\dot f}{M}_y=S (\dot f_y+[\bar Q,f_y]+\bar Q_{\bar z})S^{-1},\quad \rig{\dot f}{M}_z=S (\dot f_z+[\bar Q,f_z]-\bar Q_{\bar y})S^{-1}
$$
where $\bar Q$ is the same linear combination as $Q$ constructed not from components of $F$ function but from the corresponding derivatives of $f$ function on evolution parameter.
Indeed calculations in both cases are the same except of the last step: in the last case we 
have $\dot f$ in the first one - $f'\equiv F$.
After reducing of the second system from the first one the result will be that
transformed value $\rig{\dot f}{M}-\rig{F}{M}$ is presented as linear combinations of non transformed ones. Thus equations of hierarchy are invariant with respect to discrete transformation of maximal root of self dual Yang-Mills system.

\section{Outlook}

The main result of the present paper is the explicit form for raising Hamiltonian operator of symmetry equation of Yang Mills system. This allow to construct SDYM hierarchy, each system of which is invariant with respect to discrete transformation of the maximal root of the semi simple algebra. By the way it is solved the problem of construction all hierarchies of multi component integrable systems, which can be obtained from self dual system by reduction procedure.
Particular the hierarchy of n-waves in the case of arbitrary semi simple algebra \cite{l 2008}
with using the results of the present paper may obtained in more economical technique way.
The program of considering all integrable systems as some reducing of self dual one was proposed in the known paper of R.Ward.

Of course solution of the problem (\ref{MN}) in the most part is the happy guess. But there exist direct method, which allow by logical operations obtain the form of the raising Hamiltonian operator directly from equations of discrete transformation for solution of symmetry equation. In this connection see \cite{LY}. We hope that some body will be able to solve this problem by this way. And the last explicit form of soliton like solutions for SDYM hierarchy may be obtained as generalization of the same results in the case of semi simple algebras of the second rank \cite{4},\cite{7}.


\begin{thebibliography}{9}

\bibitem{OVS} L.Ovsiannikov, {\it Group Analysis of Differential Equations, Academic Press 1982}
\bibitem{L5}A.N.Leznov  {\it Discrete and Backlund (!) transformations of SDYM system. 
e-Print: math-ph/0504004} 
\bibitem{Lez 88} A.N.Leznov {\it Theor.Math.Phys.73:1233,1988, Teor.Mat.Fiz.73:302-307,1987.}
\bibitem{Y} C.N.Yang {\it Phys.Rev.Lett. v 38, 1337 (1997)}.
\bibitem{l 2008} A.N. Leznov {\it Two Poisson structures invariant with respect to discrete transformation in the case of arbitrary semi-simple algebras.e-Print: arXiv:0801.2541 [hep-lat]}
\bibitem{LY} A.N.Leznov and E.A.Yusbashjan {\it Integrable mappings for non commutative objects.
Published in Rept.Math.Phys.43:207-214,1999. e-Print: hep-th/9609024}
\bibitem{4} A.N.Leznov G.R.Toker and R.Torres-Cordoba {\it Nonlinear Mathematical Physics,
v, N2(2007) 238-249}
\bibitem{7} A.N.Leznov  {\it Multisoliton solution of 3,4,6-th waves problems connected with semi-simple algebras of the second rank $A_2,B_2\simeq C_2,G_2$  math-ph/0703063 v.1}

\end{thebibliography}
\end{document}